\documentclass[twocolumn,openany,openright,amsmath,amssymb,superscriptaddress,prb]{revtex4-2}
\usepackage[utf8]{inputenc}
\usepackage[T1]{fontenc}
\usepackage{float}
\usepackage{latexsym} 
\usepackage{amsmath,amsthm}
\usepackage{ifpdf}
\usepackage{epstopdf}
\usepackage{siunitx}
\usepackage{soul}
\usepackage{dcolumn}
\usepackage{bm}
\usepackage{braket}
\usepackage{wrapfig}
\usepackage{comment}
\usepackage[abs]{overpic}
\usepackage{graphicx,graphics}
\usepackage{dcolumn}
\usepackage{wasysym}
\usepackage{latexsym,verbatim}
\usepackage{color}
\usepackage[framemethod=default]{mdframed}
\usepackage[breaklinks=true,colorlinks,citecolor=blue,linkcolor=blue,urlcolor=blue]{hyperref}
\usepackage[makeroom]{cancel}
\usepackage{fancybox}
 \usepackage{tikz-feynman}
 \usepackage{lmodern}

\newcommand {\Pdert}{\frac {\partial}{\partial t}}

\newcommand{\tr}[1]{\textrm{Tr}\left( #1 \right)}

\newcommand{\eps}{\epsilon}

\newcommand{\sx}{\sigma_{\rm x}}
\newcommand{\sy}{\sigma_{\rm y}}
\newcommand{\sz}{\sigma_{\rm z}}

\newcommand{\kk}{\mathbf{k}}

\newcommand{\qq}{\mathbf{q}}

\newcommand{\EE}{\mathbf{E}}
\newcommand{\A}{\mathbf{A}}

\begin{document}
\def \k{\bm k}
\def \l{\bm l}
\def \i{\bm i}
\def \p{\bm p}
\def \ppi{\bm \pi}
\def \r{\bm r}
\def \s{\bm s}
\def \P{\bm P}
\def \R{\bm R}
\def \qq{\bm q}
\def \D{\bm D}
\def \beq{\begin{equation}}
\def \eeq{\end{equation}}
\def \beal{\begin{aligned}}
\def \eal{\end{aligned}}
\def \bes{\begin{split}}
\def \ees{\end{split}}
\def \besu{\begin{subequations}}
\def \esu{\end{subequations}}
\def \g{\gamma}
\def \G{\Gamma}
\def \ac{\alpha_c}
\def \barr{\begin{eqnarray}}
\def \earr{\end{eqnarray}}
\title{Enhanced third harmonic response of the \texorpdfstring{PtTe$_2$}{PtTe2} transition metal dichalcogenide}
\author{Leone Di Mauro Villari}
\email{leone.dimaurovillari@univaq.it}
\affiliation{Department of Physical and Chemical Sciences, University of L'Aquila, Via Vetoio, 67100 L'Aquila, Italy}
\author{Simone Grillo}
\affiliation{Department of Physics, University of Rome Tor Vergata and INFN, Via della Ricerca Scientifica 1, 00133 Roma, Italy}
\affiliation{Max Planck Institute for the Structure and Dynamics of Matter (MPSD) and Center for Free-Electron Laser Science (CFEL), 22761 Hamburg, Germany}
\author{Olivia Pulci}
\affiliation{Department of Physics, University of Rome Tor Vergata and INFN, Via della Ricerca Scientifica 1, 00133 Roma, Italy}
\author{Salvatore Macis}
\affiliation{Department of Physics,  Sapienza University of Rome, Piazzale Aldo Moro, 00133 Roma, Italy}
\author{Stefano Lupi}
\affiliation{Department of Physics,  Sapienza University of Rome, Piazzale Aldo Moro, 00133 Roma, Italy}
\author{Andrea Marini}
\email{andrea.marini@univaq.it}
\affiliation{Department of Physical and Chemical Sciences, University of L'Aquila, Via Vetoio, 67100 L'Aquila, Italy}
\affiliation{CNR-SPIN, c/o Department of Physical and Chemical Sciences, Via Vetoio, Coppito (L'Aquila) 67100, Italy}
\begin{abstract}
We investigate the third harmonic response of platinum ditelluride (PtTe$_2$), a Dirac semimetal belonging to the transition metal dichalcogenides  class. Due to its topological properties, this material has drawn a lot of attention, particularly because it hosts type-II (super-critically tilted) Dirac fermions in the $\rm A-\Gamma-\rm A$ high symmetry direction. Adopting a low-energy model fitted directly from density functional theory band structure simulations, we calculate analytically the nonlinear conductivity. We observe that  third-order optical nonlinearities are efficiently modulated by the cones tilting, which produces a significant enhancement of the nonlinear susceptibility. Our results, besides shedding light on topological transitions of platinum ditelluride, are relevant for future nanophotonic devices exploiting the tunable nonlinear properties of type-II Dirac fermions.
\end{abstract}
\maketitle
\section{Introduction}
Relativistic massless and massive fermions can be probed with high-energy physics experiments, but also appear as low-energy quasiparticle excitations in condensed matter systems, where their massless character is typically protected by crystal symmetries. In the twenty years since the discovery of graphene, Dirac and Weyl fermions (DFs-WFs) in two and three dimensions have been the subject of a remarkable amount of research in condensed matter physics \cite{ng,neto,nb,wb,ws,ww}. The class of materials showing this low-energy behavior encompasses semimetals \cite{tsa}, transition metal dichalcogenides (TMDs) \cite{xl1}, and topological insulators \cite{hak}.  Beyond standard (type-I) D-WFs new types of low-energy excitation were proposed \cite{Soluyanov2015,Huang2016}. These  quasiparticles, named type-II D-WFs, break Lorentz invariance, which is not a necessary constraint in condensed matter physics. Differently from conventional D-WFs which have standard point-like Fermi surfaces, type-II fermions emerge at the boundary between electron and hole pockets. As a consequence, instead of  straight Dirac or Weyl cones (DCs-WCs), the low-energy spectrum is given by a strongly tilted cone. Dirac and Weyl type II quasiparticles are distinguished as in the standard case: the former is symmetry protected whilst the latter ought to break either inversion or time reversal symmetry. Shortly after their theoretical prediction, type-II fermions were experimentally observed in materials such as molybdenum ditelluride (MoTe$_2$) \cite{dwd}, tungsten disulfide (WSe$_2$) \cite{wyh} and the platinum based PtX$_2$ class of TMDs (X=Te,Se,Bi) \cite{Yan2017}. Beyond fundamental science type-II fermions are particularly relevant in the field of topological materials, in fact type-I and type-II fermions can be regarded as two topological phases as the Fermi surface changes shape. The transition between the two types of fermions is called Lifshitz topological transition \cite{Soluyanov2015,Volovik2017}. More recently, the relation between topological and optical properties has started to attract attention \cite{CarvalhoMariniBiancalana2018,Ikeda2020,ZuberZhang2021,DiMauroVillariPrincipi2022}. In particular optical probing can be used to detect some signature of Lifshitz transitions \cite{Tan2022,Sie2019UltrafastSymmetrySwitch,Heide2022ProbingTPT_HHG,Bauer2022OpticallySensingTPT,Morimoto2016TopologicalNLO} which notabbly can be induced by pressure \cite{DiPietroMitrano2018}. Moreover it has been shown that the tilting of D-WCs significantly enhances  the even harmonics response in non centrosymmetric media \cite{Tamashevich2022PRB105_195102,Tamashevich2023PRB107_245425}. 

In this paper we investigate the tensorial third harmonic response of platinum ditelluride (PtTe$_2$), a $\mathcal{I}$ and $\mathcal{T}$ symmetric type-II Dirac semimetal, which has been hitherto overlooked, while its linear response is well characterized \cite{Macis2024}. Generally, crystalline third-order effects are difficult to be accounted for due to the significant complications in computing four-rank  tensors. In the case of tilted fermions, such complications are even more pronounced due to Fermi velocity anisotropy (as it is for the PtX$_2$ class of materials). The reason behind these difficulties is that the Fermi surface has no more spherical symmetry but is rather elliptic or hyperbolic. For our calculation, we start from density functional theory electronic band structure simulations, to derive a minimal model for the low-energy bands around the nodal (type-II) points in the $\rm A-\Gamma-\rm A$ high symmetry direction. We then use finite temperature diagrammatic perturbation theory to evaluate the third harmonic response. This method simplifies the calculation of high order response functions and makes the inclusion of many-body effects considerably more straightforward. We find that PtTe$_2$ has a significantly strong THG signal in the terahertz (THz) regime along the tilting direction. In fact, the THG signal from a tilted Dirac fermion is one order of magnitude larger than the signal from an untilted one. \\
 
The paper is organised as follows. In section II we compute the minimal two band model starting from the DFT electronic band structure results.  In section III, we develop the diagrammatic theory of the optical Dirac-Bloch equations. In section IV, we show the results for the third harmonic conductivity tensor.

\begin{figure*}[t]  
  \centering
  \begin{minipage}[t]{0.32\textwidth}
    \centering
    \includegraphics[width=\linewidth]{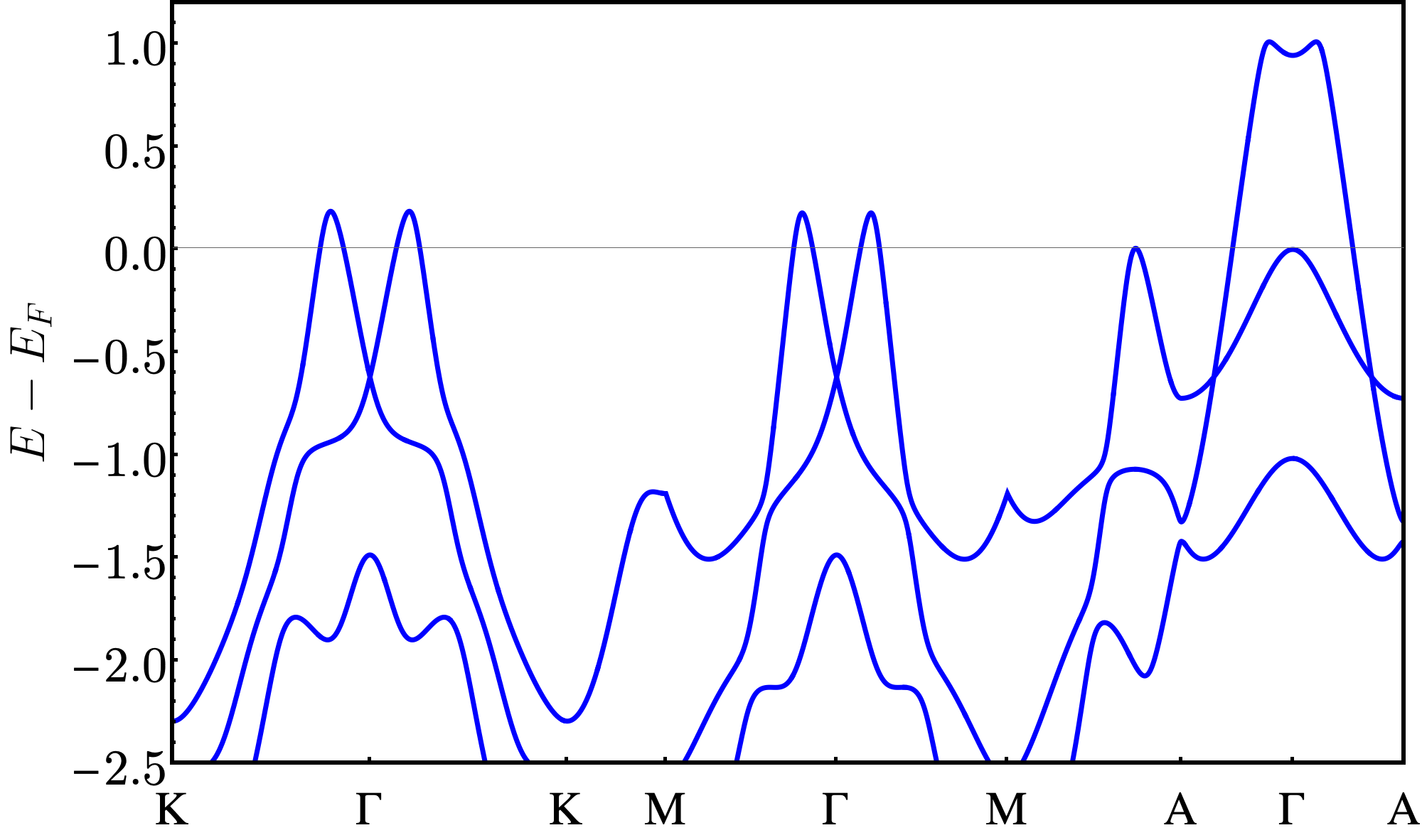}
      \put(-150,106){\textbf{(a)}} 
  \end{minipage}
\hfill
 \begin{minipage}[t]{0.32\textwidth}
    \centering
    \includegraphics[width=\linewidth]{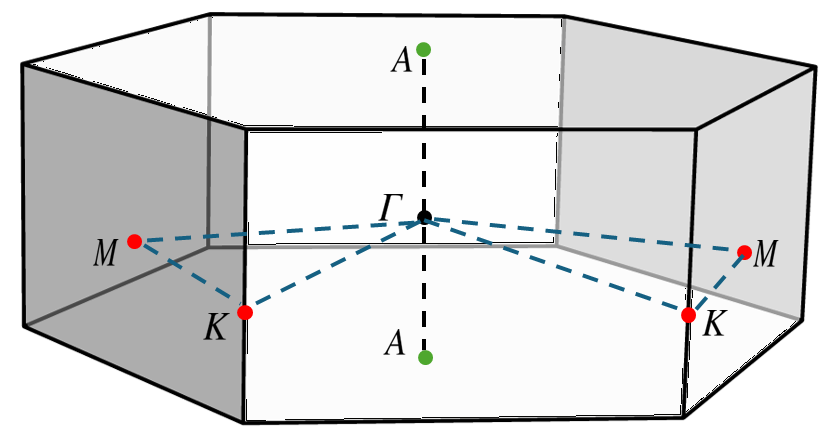}
      \put(-150,106){\textbf{(b)}} 
  \end{minipage}
  \hfill
  \begin{minipage}[t]{0.32\textwidth}
    \centering
    \includegraphics[width=\linewidth]{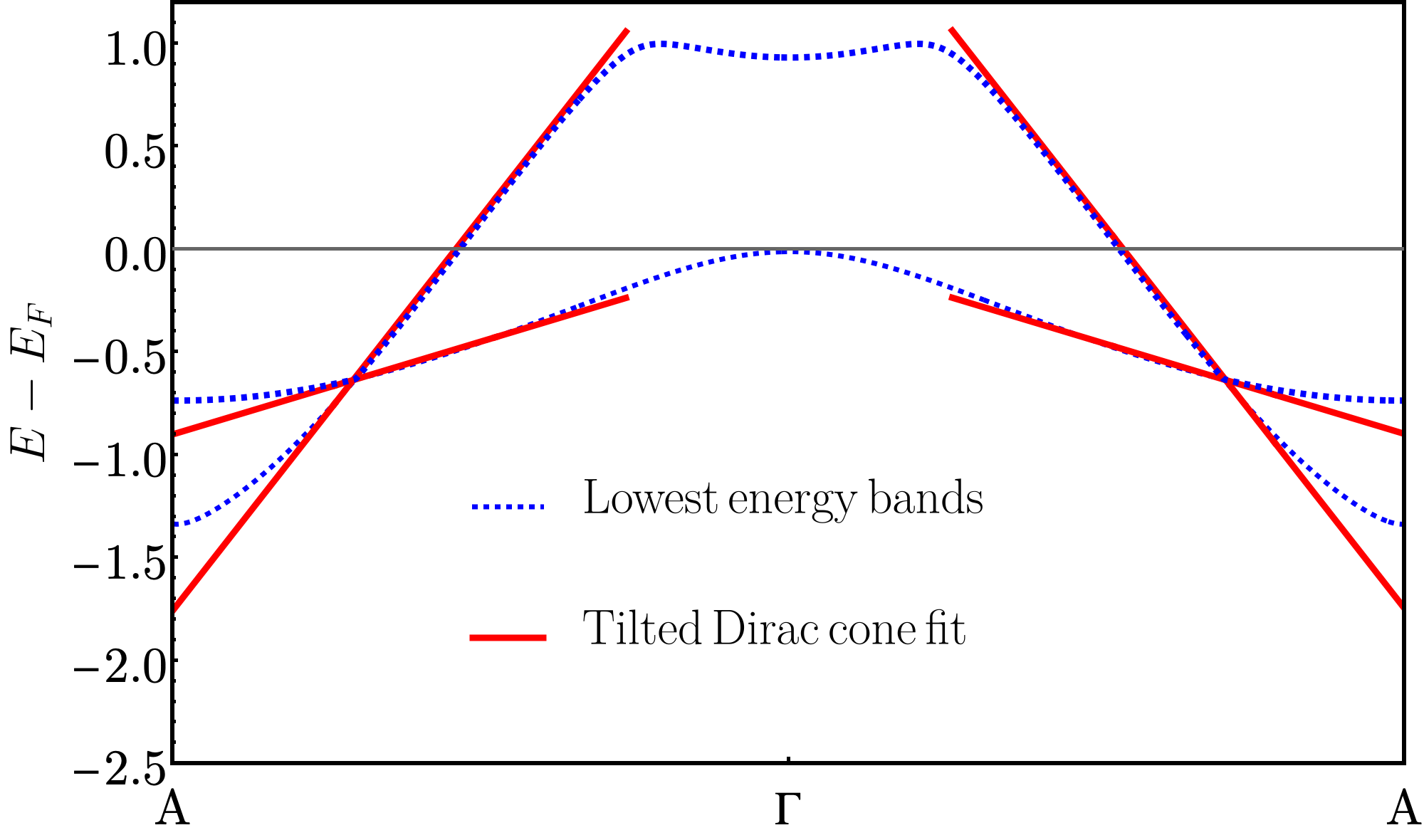}
        \put(-150,106){\textbf{(c)}} 
  \end{minipage}
  \caption{{\bf (a)} Bulk 1T-PtTe$_2$  electronic band structure, with SOC included, calculated with Grimme’s PBE-DFT-D2 functionals, with high-symmetry directions in the hexagonal Brillouin zone at $k_z = 0.35(4)c^*$, where $c^* = 2\pi/c$ (we use $\Gamma$ as a reference as $k_z = 0$). Energy rescaled with respect to the Fermi energy. It shows two type-I Dirac points in the $\rm K$-$\Gamma$-$\rm K$ and $\rm M$-$\Gamma$-$\rm M$ high symmetry direction, and two type-II points in the $\rm A$-$\Gamma$-$\rm A$ direction. {\bf (b)} Wave-vector path in the hexagonal Brillouin zone. {\bf(c)} Lowest energy bands and fit with the low momentum expansion around the type-II nodes. }
  \label{fig1}
\end{figure*}

\section{Low-energy model from DFT band structure}
In this section we briefly review the electronic properties of 1T-PtTe$_2$, which is a metal/semimetal in the bulk TMDs family. The crystal structure has trigonal symmetry and a three-atom unit cell, consisting of two Te atoms and one Pt atom.

The electronic band structure was calculated within DFT using the Quantum ESPRESSO package \cite{Giannozzi2009,Giannozzi2017}, employing a norm-conserving, fully relativistic pseudopotential from the PseudoDojo repository \cite{vanSetten2018} including spin-orbit coupling and semicore corrections, together with a generalized gradient approximation (GGA), according to Perdew, Burke and Ernzerhof (PBE) \cite{Perdew1996}, for the exchange-correlation functional. A kinetic energy cutoff of 100 Ry and a $12\times12\times12$ Monkhorst–Pack \cite{Monkhorst1976} $k$-point mesh were adopted after careful convergence tests. van der Waals interactions were taken into account using Grimme’s PBE-DFT-D2 correction scheme.
 We then have derived the effective two-band model by fitting the lowest energy bands of the DFT band structure which we show in figure \ref{fig1}a, with the following model-Hamiltonian (see Fig. \ref{fig1}c)
\begin{equation} \label{eq1}
H= a_0 \sigma_0 k_z +\sum_{i=(x,y,z)} a_i \sigma_i k_i,
\end{equation}
where $a_i$ are fitting coefficients, $\sigma_i$ the Pauli matrices plus the identity (indicated with $\sigma_0$) and $k_i$ the electron momentum. The fit gives the following values for the coefficients, $a_0=-6.25 \, \rm{eV}\,\mathring {\mathrm A}$, $a_x=2.76 \, \rm{eV}\,\mathring {\mathrm A}$, $a_y=-a_x$ and $a_z=-3.86 \, \rm{eV}\,\mathring {\mathrm A}$. We can now rewrite the two bands Hamiltonian in a more convenient form by introducing the dimensionless tilting parameter $\tau=a_0/a_z$ and Fermi velocity anisotropy $(\eta_x=a_x/a_z,\eta_y=a_y/a_z)$, hence
\beq \label{eq2}
H= \eta_x \sx q_x + \eta_y \sy q_y + (\tau \sigma_0 + \sz)q_z,
\eeq
where the rescaled "momentum" $q_i=a_z k_i$ has the dimensions of an energy. With the Hamiltonian written in this form we can distinguish three cases $\tau<1$ (tilted type-I DFs) $\tau=1$ (critically tilted), $\tau>1$ (supercritically tilted type-II DFs). In the case of PtTe$_2$, the tilting parameter is $\tau=1.54$ hence it shows type-II DFs as we can clearly see in Figs. \ref{fig1}a,b.

The Hamiltonian in Eq. (\ref{eq2}) can be easily diagonalized, giving 
\beq
\begin{aligned}
\mathbf{u}_{\qq}^\lambda = \frac{g_{\qq}}{f_{\qq}\sqrt{2w_{\qq,\lambda}}}\begin{pmatrix}
 \frac{f_{\qq} w_{\qq,\lambda}}{g_{\qq}}e^{-i\varphi_{\qq}}, \\
1
\end{pmatrix}\\
\varepsilon^\xi_{\qq,\lambda}= \xi q_z + \lambda \sqrt{\eta_x^2 q_x^2 + \eta_y^2 q_y^2 + q_z^2  },
\end{aligned}
\eeq
where $\lambda=\pm1$ is the band index, $\xi=\pm 1$ is a pseudo-spin index which runs on the two type-II cones in the $\rm A-\Gamma-\rm A$ direction, while $g_{\qq}=\sqrt{\eta_x^2 q_x^2 + \eta_y^2 q_y^2}$, $f_{\qq}=\sqrt{\eta_x^2 q_x^2 + \eta_y^2 q_y^2+q_z^2}$ and $w_{\qq,\lambda}=1+\lambda \,  q_z/f_{\qq}$.

\section{Feynman diagrammatic expansion of the Bloch equations}
This section gives a short overview of the calculation of the nonlinear conductivity starting from the non-interacting semiconductor Bloch equations or optical Bloch equations (OBEs) in  a mixed velocity/length gauge \cite{Cheng_2014}, which can be represented as a diagrammatic expansion.  We choose a time-dependent frame which moves with the vector potential. Hence, the OBEs describe the evolution of the density matrix as (for the sake of conciseness, in this section we take $\hbar=c=e=1$, unless specified)
\beq
i \Pdert \rho_{\kk} = [H_{\kk-\A(t)}[\EE(t)],\rho_{\kk}],
\eeq
where $\A(t)$ is the electromagnetic (EM) vector potential and $\EE(t)=-\dot\A(t)$ is the electric field. Taking the expectation values over the bands eigenstates gives 
\beq \label{SBE}
i\Pdert \rho^{\lambda,\lambda'}_{\kk} = (\varepsilon_{\kk+ \A(t)})\rho_{\kk}^{\lambda,\lambda'} - \EE(t)\cdot \mathcal{A}^{\lambda,\lambda'}_{\kk}\rho_{\kk}^{\lambda,\lambda'} ,
\eeq

\begin{figure}[t]
    \centering
    \includegraphics[width=\linewidth]{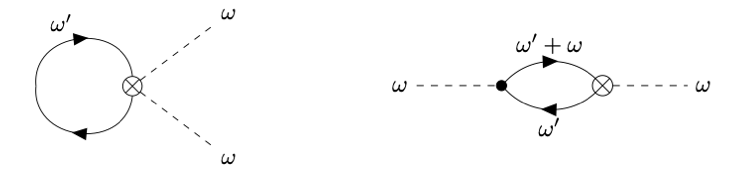}
    \caption{First order conductivity diagrams.}
    \label{fig2}
\end{figure}

\noindent where the matrix $\mathbf{\mathcal{A}}^{\lambda,\lambda'}_{\kk} = \bra{\mathbf u^{\lambda}_{\kk}}\nabla_{\kk} \ket{\mathbf{u}^{\lambda'}_{\kk}}$ is the non-abelian Berry connection and $\lambda$ is a band index. This equation accounts for both the interband and intraband dynamics of the carriers. As we are interested in the nonlinear response, we express the full current as 
\beq
\mathbf{J}(t)=\int \frac{d\kk}{(2\pi)^3} \tr{\mathbf{j}_{\kk-\A(t)}^{\lambda,\lambda'} \rho^{\lambda,\lambda'}_{\kk}(t)},
\eeq
where the $\mathbf{j}^{\lambda,\lambda'}_{\kk}= \bra{\mathbf u^{\lambda}_{\kk}}\nabla_{\kk} H_{\kk} \ket{\mathbf{u}^{\lambda'}_{\kk}}$ is the microscopic current operator. The current can be expanded perturbatively as \cite{boyd}
\beq
\begin{split}
\mathbf{J}(t) = \sum_n \int \frac{\Pi_{i=1}^n d\omega_i}{2\pi} \hat \sigma^{(n)}(\omega_\Sigma;\omega_1,\dots,\omega_n) e^{-i\omega_\Sigma t} \bigotimes_{i=1}^n\EE_{\omega_i},
\end{split}
\eeq
where $\omega_i$ are external EM frequencies, $\omega_\Sigma=\omega_1+\dots+\omega_n$, $\hat\sigma^{(n)}$ the $n$-th order conductivity tensor and $\EE_{\omega_1}\otimes \EE_{\omega_2}=E^a_{\omega_1} E^b_{\omega_2}$, with $a,b=(x,y,z)$ and $\EE_\omega$ the electric field Fourier transform. The nonlinear orders of the conductivity response functions can be found explicitly from the perturbative solution of the OBEs. The procedure is straightforward, but rather lengthy above all at high order. To this end a Feynman diagrammatic expansion is rather helpful. Depending on the gauge used and the response function considered, Feynman expansions have slightly different rules \cite{Parker2019_DiagrammaticNonlinear}, even though they all give the same results if diagrams are evaluated carefully. We summarise here the rules for our specific case (from here we reintroduce physical units): 
\begin{itemize}
    \item draw electrons closed solid lines (loops) in all possible topologies allowed by the expansion order ($n$), i.e., with a maximum of $n+1$ vertices. This gives $2^n$ diagrams at each order;
    \item associate a frequency  to each solid or interaction line. Conserve frequency at each vertex;
\item associate to each solid line a free electron propagator 
\[  G_\lambda(\kk,\Omega_j)= \frac{1}{\Omega_j - \varepsilon_{\lambda,\kk}},\]
where $\Omega_j$ are combinations of external photon frequencies and internal electron frequencies fixed by conservation of frequency;
    \item  draw $n+1$ external dashed photon lines to be "attached" to electron vertices;
     \item associate a frequency $\omega_i$ to each photon line attached to an input vertex and a frequency $\omega_{\Sigma}$ to photon lines attached to an output vertex;
    \item identify input and output interaction vertices, each diagram has one output vertex (with one output photon and a maximum of $n$ input photons attached) and a maximum of $n$ input vertices (with a maximum of $n$ input photons attached);
    \item associate the appropriate current matrix element at each input and output vertex. An $m$-frequencies input vertex is given by
    \[
    \mathcal J^{(m)\alpha_1,\dots,\alpha_m}_{\lambda\lambda',\kk} = e^m\frac{\partial^{m-1}}{\partial k_{\alpha_1} \dots \partial k_{\alpha_{m-1}}} \Biggl(\frac{\partial \varepsilon^\lambda_{\kk}}{\partial k_{\alpha_m}} + m\,  i \omega \mathcal A^{\lambda\lambda'}_{\kk}\Biggl)
    ,\]
    while an $l$-frequency output vertex is given by 
    \[
    J^{(l)\alpha_1,\dots,\alpha_l}_{\lambda\lambda',\kk} = e^l \frac{\partial^{l-1}}{\partial k_{\alpha_1} \dots \partial k_{\alpha_{l-1}}}\bra{u^\lambda_{\kk}} \frac{\partial H_{\kk}}{\partial k_{\alpha_l} }
   \ket{u^{\lambda'}_{\kk}}, \]
   with $m+l=n+1$;
    \item integrate over the internal electron wavevector and frequencies with measures $\int d^dk/(2\pi)^d$ and $\int d\omega/(2\pi) $;
    \item multiply everything by a factor $\frac{i^n}{\hbar^{n+1} \prod_{i=1}^n\omega_i}$.
\end{itemize}

Equipped with these rules, we can draw all diagrams at each order and give the respective analytical expression. We consider the first order, as it is simpler and instructive, and the third-order, which is the one we are interested in. The second order (along with all even orders) is vanishing as the material is centrosymmetric.

\begin{figure}[b]
    \centering
    \includegraphics[width=\linewidth]{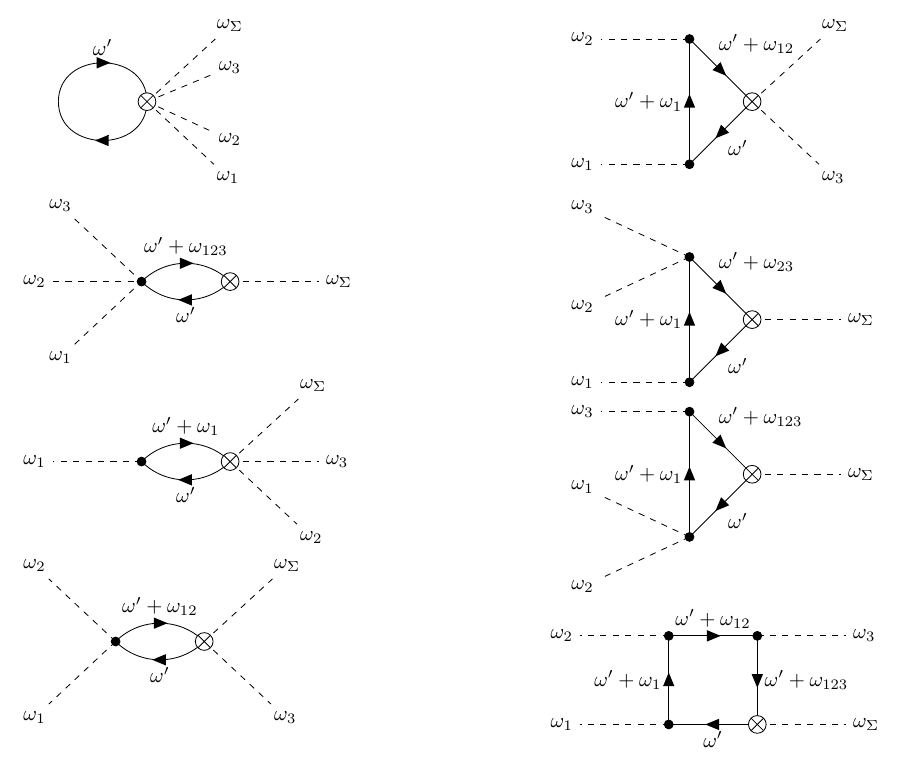}
    \caption{Third-order conductivity diagrams.}
    \label{fig3}
\end{figure}

\begin{figure*}[t]
  \centering

  \begin{minipage}[t]{0.48\textwidth}
    \centering
    \begin{overpic}[width=\linewidth]{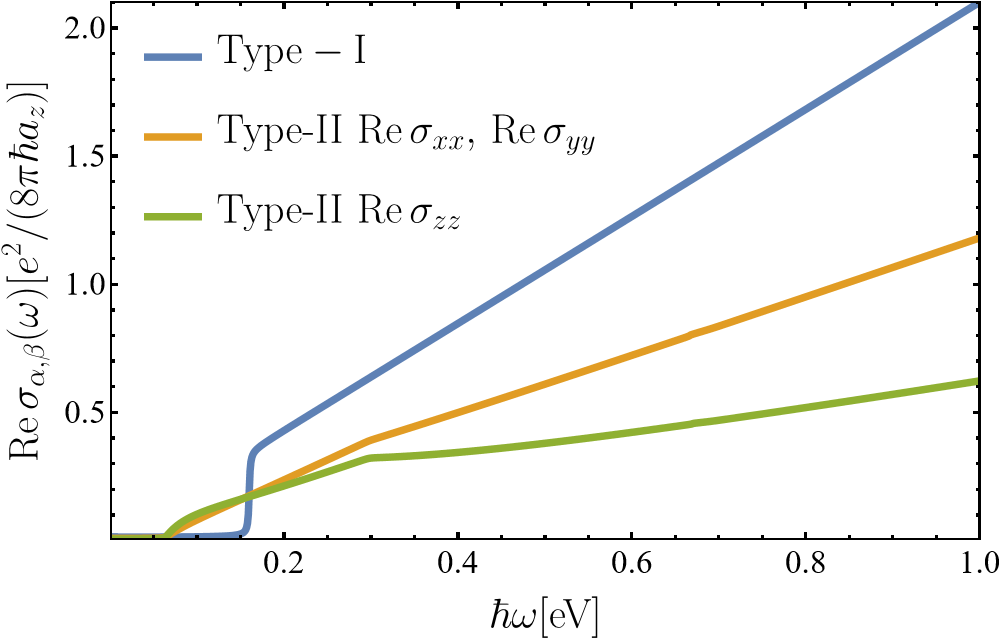}
      \put(2,166){\textbf{(a)}}
    \end{overpic}
  \end{minipage}\hfill
  \begin{minipage}[t]{0.48\textwidth}
    \centering
    \begin{overpic}[width=\linewidth]{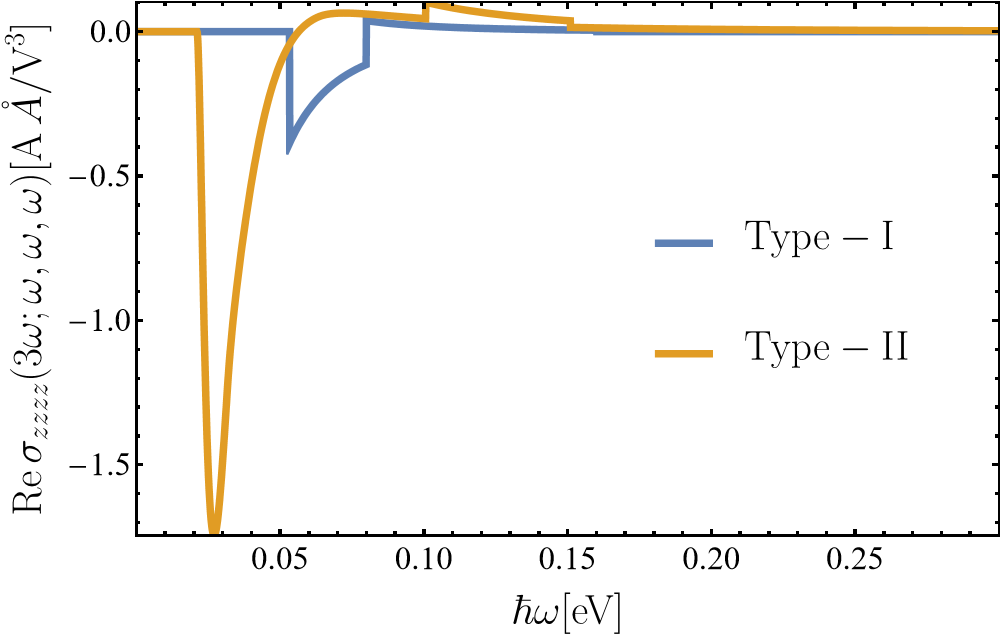}
      \put(2,166){\textbf{(b)}}
    \end{overpic}
  \end{minipage}

  \vspace{0.8em}

  \begin{minipage}[t]{0.48\textwidth}
    \centering
    \begin{overpic}[width=\linewidth]{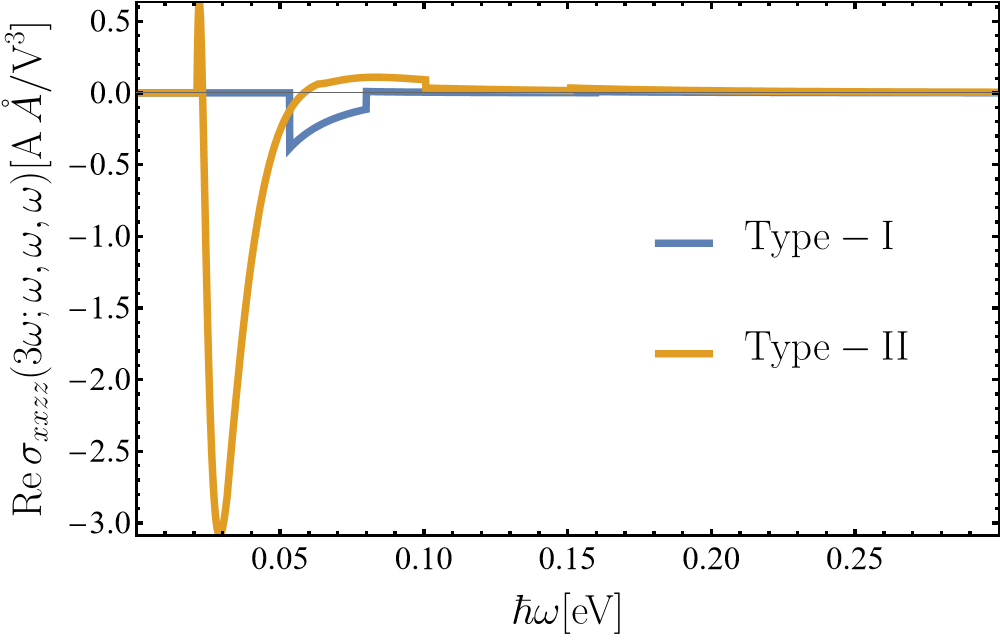}
      \put(2,166){\textbf{(c)}}
    \end{overpic}
  \end{minipage}\hfill
  \begin{minipage}[t]{0.475\textwidth}
    \centering
    \begin{overpic}[width=\linewidth]{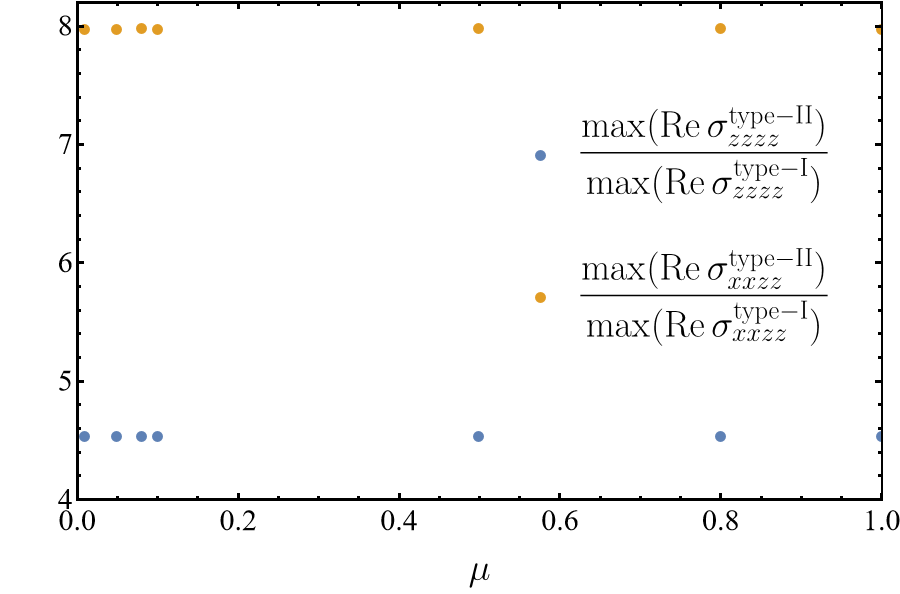}
      \put(2,166){\textbf{(d)}}
    \end{overpic}
  \end{minipage}

  \caption{Real part of the first order and THG nonlinear conductivity tensor at $T=0$. \textbf{(a)} first order response for PtTe$_2$ (type-II, $xx=yy$ and $zz$-directions) and an ideal massless type-I Dirac cone $(xx=yy=zz).$ \textbf{(b)} $\sigma^{(3)}_{zzzz}(\omega)$ and \textbf{(c)} $\sigma^{(3)}_{xxzz}(\omega)$ for PtTe$_2$ (type-II) and an ideal massless type-I Dirac cone. \textbf{(d)} THG enhancement, defined as the ratio between the maximum conductivity of the type-II and type-I nodes, as a function of chemical potential.}
  \label{fig4}
\end{figure*}

The first diagram gives the intraband contribution to the conductivity while the second gives the interband contribution. According to the rules, such contributions can be represented as in Fig. \ref{fig2} and their analytical expressions read

\beq
\begin{split}
\sigma^{(1)}_{\alpha \beta}(\omega;\omega)=&\frac{i}{\hbar^2 \omega} \int \frac{d^3 k}{(2\pi)^3} \int \frac{d\omega'}{2\pi}\Biggl[ \sum_\lambda  J_{\lambda,\kk}^{\alpha \beta} G^{\lambda}_{\kk}(\omega') d\omega' +  \\& +  \sum_{\lambda \lambda'}  
 \mathcal J^\alpha_{\lambda\lambda',\kk} G^{\lambda'}_{\kk}(\omega+\omega')J^\beta_{\lambda\lambda',\kk}G^{\lambda}_{\kk}(\omega')\Biggl],
\end{split}
\eeq
where the integration over internal frequencies can be performed using the standard technique of the Matsubara summation  \cite{Giuliani_Vignale_2005}, while the integration over the momentum can be arbitrarily difficult and depend on the system under consideration. At third-order there are $2^3=8$ different diagrams which we show in Fig. \ref{fig3}. The analytical expression of the third-order conductivity is rather cumbersome and we refer to appendix \ref{AppA} where we explain how to compute it explicitly.

\section{Topological third harmonic response}

In this section we show the result of the diagrammatic calculation for the  third harmonic response, for which $\omega_1=\omega_2=\omega_3=\omega$. PtTe$_2$ belongs to the $D_{3d}$ symmetry group, hence the third-order conductivity tensor has 14 nonzero independent elements. Qualitatively their behaviour is rather similar, even though the enhancement due to the presence of type-II Dirac cones is more evident in the $k_z$-tilting direction. For this reason we shall show the results for the $\sigma_{zzzz}$ and $\sigma_{xxzz}=\sigma_{yyzz}$ terms.
The details of the diagrammatic calculations, for the bubble and triangle diagrams are shown in  appendix \ref{AppA}. Before detailing the THG, we review in figure \ref{fig4}a the first order response comparing type-I and type-II DCs \cite{Carbotte2016DiracConeTilt}. We can observe first that for type-I DCs $\sigma_{xx}=\sigma_{yy}=\sigma_{zz}$ while for type-II $\sigma_{xx}=\sigma_{yy}\neq \sigma_{zz}$. Secondly the resonances and analyticity properties are changed, for type-I the conductivty is zero before $\hbar \omega=2\mu$ (with $\mu$ being the chemical potential) and is linear in energy for $\hbar \omega>2\mu$, for the case of type-II to the onset of the interband transitions is at $\hbar \omega=2\mu/(1+\tau)$  and they show different quasi-linear behaviors for $2\mu/(\tau+1)<\hbar \omega <2\mu/(\tau-1)$ and $\hbar \omega> 2\mu/(\tau-1)$ . The results are shown in Figs. \ref{fig4}b-d. In Figs.  \ref{fig4}b,c we show the $\sigma_{zzzz}$ and $\sigma_{xxzz}$ for the THG process, both for a type-I and a type-II DC with chemical potential $\mu=0.08 \, \rm eV$. We can see that  type-II cones show a  significantly larger signal for both the tensor terms, at low-energy (with respect to the nodal point). This is due to the fact that more $k$-states become available for transitions at low-energy, on the contrary at higher energy less states are available in the tilted case (this is also clearly visible in the first-order response). As we can observe from Fig. \ref{fig4}d, the THG signal is enhanced by $4.5$ times for the $zzzz$-tensor element and by $8$-times for the $xxzz$-tensor element and is independent from the doping (in the considered range and within the validity of the low-energy model). Considering that third-order effects are generally very low, such an increase is remarkable and makes Pt based TMDs excellent platforms for nonlinear optical frequency conversion, as recently confirmed experimentally \cite{Zhu2023Ultrastrong}. 
Moreover, as it can be seen from Figs. \ref{fig4}a and b, the resonance structure of the nonlinear response is affected by the topological transition (i.e. the presence of type-II nodes). The expressions of the real part of THG conductivity can be compactly written as 
\begin{widetext}
\begin{equation}
\begin{aligned}
    &\textrm{Re}\,\sigma^{\rm tyI}_{\alpha,\beta,\g,\delta}(3 \omega;\omega,\omega,\omega)= f^{(1)}_{\alpha,\beta,\g,\delta}(\omega) \Theta \Bigl(\frac{2}3\mu-\hbar \omega\Bigl) + f^{(2)}_{\alpha,\beta,\g,\delta}(\omega) \Theta (\mu-\hbar \omega) + f^{(3)}_{\alpha,\beta,\g,\delta}(\omega) \Theta (2\mu-\hbar \omega) \\
   &\textrm{Re}\,\sigma^{\rm tyII}_{\alpha,\beta,\g,\delta}(3 \omega;\omega,\omega,\omega)= \sum_{m=\pm 1} \Bigl[g^{(1),m}_{\alpha,\beta,\g,\delta}(\omega,\tau) \Theta \Bigl(\frac{2\mu}{3(\tau + m)}-\hbar \omega\Bigl) + g^{(2),m}_{\alpha,\beta,\g,\delta}(\omega,\tau) \Theta \Bigl(\frac{\mu}{\tau + m}-\hbar \omega\Bigl) + g^{(3),m}_{\alpha,\beta,\g,\delta}(\omega,\tau) \Theta \Bigl(\frac{2\mu}{\tau + m}-\hbar \omega\Bigl)\Bigl]
    \end{aligned}
\end{equation}
\end{widetext}
The multiplying functions $(f,g)$ are in general given by rather complicated integrals which are read off from the diagrammatic expansion (see appendix \ref{AppA}). The presence of the tilting modifies the position of the resonances in energy space, which are given by the arguments of the Heaviside $\theta$-functions. In the case of type-I DCs (blue curve in Fig. \ref{fig4}a,b) there are three resonances, located at $\hbar \omega= (2/3\mu,\mu,2\mu)$ while for the case of type-II, due to tilting, there are six resonances $\hbar \omega= \{2\mu/[3(\tau \pm 1),\mu/(\tau\pm 1),2\mu/(\tau\pm1)\}$. Thanks to the matching between Heaviside functions, which is tuned by the tilting parameter $\tau$, the response of PtTe$_2$ type-II DCs is smooth in the interval $2\mu/[3(\tau + 1)\leq\hbar \omega <2\mu/[3(\tau - 1)$. It shows, instead, $\Theta$-singularities at $\hbar \omega=\{2 \mu/ [3(\tau + 1),2 \mu/ (\tau + 1),\mu/ (\tau + 1)\}$.
\section{Conclusion}

In this paper we studied the nonlinear optical response of the PtX$_2$ class of TMDs. We have focussed on PtTe$_2$, but our calculations holds for all materials of the same class as they share the same electronic and symmetry properties. In section II we presented DFT calculation of the electronic band structure from which we derived a low-energy two-band model in the $\rm A-\Gamma-A$ high symmetry direction, where it hosts two symmetric type-II Dirac fermions. We then analytically diagonalized the two bands Hamiltonian. In section III we derived a Feynman diagrammatic expansion of the OBEs and listed the rules for the drawing of the diagrams at each order. Finally, in section Iv we presented the results of the fully analytical calculation of the third-order diagrams. We show a significant enhancement (up to 8 times) of the third harmonic response due to the presence of the type-II Dirac cones together with a tilted induced shifting of the resonances and a change in the analyticity properties of the response function. This suggests that Pt based TMDs are an excellent platform for nonlinear optical interactions and that nonlinear spectroscopy may be employed to get indirect evidences of topological transitions. As a conclusive remark we note that we have limited our study to the single electron approximation i.e. we have neglected many-body interactions. From a qualitative point of view this approach is rather accurate at least far from excitonic resonances \cite{HaugKoch2009QuantumTheorySemiconductors}. In perspective, to account for quantitative corrections the formalism employed in this paper  may be extended to include many body effects. 
\appendix
\section{Explicit calculation of some diagrams} \label{AppA}
In this appendix we show the explicit calculation of third harmonic diagrams. We focus on the real part, the imaginary one can either be calculated in a similar fashion or numerically using the Kramers-Kronig relations. Let us start with the bubble diagrams, the first one  describes the intraband transitions and only contributes to the imaginary part. While the remaining three bubble diagrams contribute to both the real and imaginary part. The sum of these three diagrams, after performing the Matsubara summation, reads 
\beq
\begin{split}
B_{\alpha\beta\gamma\delta}(3\epsilon,\epsilon,\epsilon,\epsilon) = -\frac{i}{2 \hbar \epsilon^3 (2\pi)^3}\sum_{n=1}^3 \sum_{\{\lambda_i\}} \\ \int d\kk \, \mathcal B^{n,\alpha\beta\gamma\delta}_{\lambda_1,\lambda_2,\kk}\frac{n^{\lambda_1,\lambda_2}_{\kk}}{n\epsilon^+ - \varepsilon^{\lambda_1,\lambda_2}_{\kk}},
\end{split}
\eeq
where $\epsilon=\hbar \omega$ and $\epsilon^+ = \hbar \omega + i \gamma$, $\mathcal B^{\alpha\beta\gamma\delta}_{n,\lambda_1,\lambda_2,\kk}$ is the vertex function for each bubble diagram. Using a Pauli-Villars-like regularization we can separate the integrand into two contributions
\beq \label{bubble}
\begin{split}
B_{\alpha\beta\gamma\delta}(3\epsilon,\epsilon,\epsilon,\epsilon)= -\frac{i  a_z}{ \hbar \epsilon^3 (2\pi)^3 \eta^2}\sum_{n=1}^3 \int_{4\pi} d\Omega \tilde {\mathcal B}^{n,\alpha\beta\gamma\delta}_{1-1,\theta,\phi}\\\int_0^\infty\,dq \Bigl[ \frac{2}n\Bigl(\frac{1}{n\epsilon^+-2q}-\frac{1}{n\epsilon^++2q}\Bigl) + \frac{1
}q\Bigl] n^{1,-1}_{q,\theta},
\end{split}
\eeq
where we have used elliptical coordinates ($q_x= q/\eta \cos \phi \sin \theta, q_y= q/\eta \sin \phi \sin \theta,q_z=q/\eta_z \cos \theta$) with $q_i=a_z k_i$ and $\eta=\eta_x=-\eta_y$, we have defined the rescaled vertex function as $\mathcal B^{n,\alpha\beta\gamma\delta}_{1-1,\theta,\phi}= \epsilon/q^3 \tilde {\mathcal B}^{n,\alpha\beta\gamma\delta}_{1-1,\theta,\phi}$ and summed over the band index $\lambda_i$. It can be shown that in the limit $\tau=0$ (massive Dirac fermions), the second term has the well-known logarithmic UV-divergence for each diagram, which cancels out under resummation of the three bubble diegrams. For tilted cones, instead, while having also a diverging contribution (vanishing under resummation) it also gives a finite imaginary contribution. Using now the Sokhotski-Plemelj formula over the real line
\beq
\frac{1}{x\pm i\gamma}= \mp i \pi \delta(x) + \mathcal P \Bigl[\frac{1}x\Bigl]
\eeq
where $\mathcal P[\cdot]$ indicates the Cauchy principal value, we can isolate the real part of the bubble conductivity
\beq
\begin{split}
\text{Re} [B_{\alpha\beta\gamma\delta}(3\epsilon,\epsilon,\epsilon,\epsilon)]= -\frac{\pi a_z}{ \hbar \epsilon^3 (2\pi)^3 \eta^2}\sum_{n=1}^3 \int_{4\pi} d\Omega \, \tilde {\mathcal B}^{n,\alpha\beta\gamma\delta}_{1-1,\theta,\phi}\\\int_0^\infty\,dq \Bigl[ \frac{2}n\Bigl(\delta(n\epsilon^+-2q)-\delta(n\epsilon^++2q)\Bigl)\Bigl] n^{1,-1}_{q,\theta},
\end{split}
\eeq
as stated in the main text this expression can be computed analytically for every non-zero tensor element even for non-zero temperature, however the expressions are cumbersome and in general not particularly instructive especially for finite temperature. To show the main feature of the result we report explicitly the $zzzz$-element for $T=0$ whose main qualitative characteristics hold for every tensor element. We get
\beq
\begin{split}
\text{Re} [B_{zzzz}(3\epsilon,\epsilon,\epsilon,\epsilon)]=\frac{a_z }{8 \pi \hbar\,\eta^2 \epsilon^3} \sum_{n=1}^3\int_{-1}^1 dw \, \tilde {\mathcal B}^{n,zzzz}_{1-1,w}\\\frac{1}n [\Theta(n\frac{\epsilon}2(1+\tau w)-\mu)- \Theta(-n\frac{\epsilon}2(1-\tau w)-\mu)],
\end{split}
\eeq
where $w=\cos \theta$, $\mu$ the chemical potential and $\Theta(x)$ the Heaviside step function. The rescaled vertex functions for the three diagrams are given by
\beq
\begin{aligned}
\tilde {\mathcal B}^{1,zzzz}_{1,-1,w} &= (1 - w^2) (6 w^2 - 2)/4 \\
\tilde {\mathcal B}^{2,zzzz}_{1,-1,w} &= 4 (1 - w^2) w^2,\\
\tilde {\mathcal B}^{3,zzzz}_{1,-1,w} &= 3 (1 - w^2) (4 w^2 - 1),
\end{aligned}
\eeq
hence we have reduced the seemingly complicated integral in equation \ref{bubble} to a sum of polynomial integrals. The analytical result for the real part of the THG bubble conductivity in the $zzzz$-direction at $T=0$ is given by
\beq
\begin{split}
\text{Re} [B_{zzzz}(3\epsilon,\epsilon,\epsilon,\epsilon)] = \frac{e^4 a_z}{8 \pi \hbar\,\eta^2 \epsilon^3}\\ \Bigl[b^1_-(\epsilon,\tau)\Theta\Bigl(\hbar \epsilon-\frac{2\mu}{3(1+\tau)}\Bigl)  + b^2_-(\epsilon,\tau) \Theta\Bigl(\hbar \epsilon-\frac{\mu}{1+\tau)}\Bigl) + \\ b^3_-(\epsilon,\tau) \Theta\Bigl(\hbar \epsilon-\frac{2\mu}{1+\tau}\Bigl) + b^1_+(\epsilon,\tau)\Theta\Bigl(\hbar \epsilon-\frac{2\mu}{3(\tau-1)}\Bigl)  + \\ b^2_+(\epsilon,\tau) \Theta\Bigl(\hbar \epsilon-\frac{\mu}{\tau-1)}\Bigl) + b^3_+(\epsilon,\tau) \Theta\Bigl(\hbar \epsilon-\frac{2\mu}{\tau-1}\Bigl)\Bigl],
\end{split}
\eeq
the $b^i_\pm(\epsilon,\tau)$ functions are defined as
\begin{widetext}
\beq
\begin{aligned}
    b^1_\pm(\epsilon,\tau)&=-\frac{2}{1215 \epsilon^5}(405 \tau ^4 \epsilon ^4 (2 \mu \pm 3 \epsilon )-75 \tau ^2 \epsilon ^2 (2 \mu  \pm 3 \epsilon
   )^3+4 (2 \mu \pm 3 \epsilon )^5 +162 \tau ^5 \epsilon ^5 ), \\
   b^2_\pm(\epsilon,\tau)&=\frac{1}{15 \epsilon^5} \left(4 \tau ^5 \epsilon ^5+15 \tau ^4 \epsilon ^4 (2 \mu \pm \epsilon )-20 \tau ^2 \epsilon ^2 (2 \mu \pm \epsilon
   )^3+9 (2 \mu \pm \epsilon )^5\right),\\
    b^3_\pm(\epsilon,\tau)&= \frac{4}{15\epsilon^5} \left(2 \tau ^5 \epsilon ^5+5 \tau ^2 \epsilon ^2 (\mu \pm \epsilon )^3-3 (\mu \pm \epsilon )^5\right)
\end{aligned}
\eeq

The triangle conductivity diagrams are instead given by the following expression
\beq
\begin{aligned}
T_{\alpha,\beta,\gamma,\delta}(3\epsilon,\epsilon,\epsilon,\epsilon)&=-\frac{i}{\hbar (2\pi \epsilon)^3}\sum_{\{\lambda_i\}} \int d\kk\, d\epsilon'\Bigl[\, G(\epsilon')\mathcal J^{\alpha}_{\lambda_1,\lambda_2,\kk} G(\epsilon'+ \epsilon_1) \mathcal J^{\beta}_{\lambda_2\lambda_3,\kk}G(\epsilon'+ \epsilon_{1,2}) J^{\gamma;\delta}_{\lambda_3\lambda_1,_{\kk}}+ \\ &\frac{1}{2!}G(\epsilon') \mathcal J^{\alpha}_{\lambda_1,\lambda_2,\kk}G(\epsilon'+ \epsilon_1) \mathcal J^{\beta;\gamma}_{\lambda_2\lambda_3,\kk}G(\epsilon'+\epsilon_{123}) J^{\delta}_{\lambda_3\lambda_1,_{\kk}} + \\ &\frac{1}{2!}G(\epsilon') \mathcal J^{\alpha;\beta}_{\lambda_1,\lambda_2,\kk}G(\epsilon'+ \epsilon_{1,2}) \mathcal J^{\gamma}_{\lambda_2\lambda_3,\kk}G(\epsilon'+\epsilon_{123}) J^{\delta}_{\lambda_3\lambda_1,_{\kk}}\Bigl].
\end{aligned}
\eeq
Performing again the Matsubara summation we get
\beq
\begin{aligned}
T_{\alpha\beta\gamma\delta}(3\epsilon,\epsilon,\epsilon,\epsilon)= -\frac{i}{\hbar (2\pi \epsilon)^3}\sum_{\{\lambda_i\}}\int d\kk \, \Biggl[ \frac{\mathcal T^{1,\alpha,\beta,\gamma,\delta}_{\lambda_1\lambda_2\lambda_3,\kk}}{2\epsilon^+ -\varepsilon^{\lambda_3\lambda_1}_{\kk}} \Biggl(\frac{n^{\lambda_1\lambda_2}_\kk}{\epsilon^+-\varepsilon^{\lambda_2\lambda_1}_\kk}+ \frac{n^{\lambda_3\lambda_2}_\kk}{\epsilon^+-\varepsilon^{\lambda_3\lambda_2}_\kk}\Biggl)+ \\\frac{\mathcal T^{2,\alpha,\beta,\gamma,\delta}_{\lambda_1\lambda_2\lambda_3,\kk}}{3\epsilon^+ -\varepsilon^{\lambda_3\lambda_1}_{\kk}} \Biggl(\frac{n^{\lambda_1\lambda_2}_\kk}{\epsilon^+-\varepsilon^{\lambda_2\lambda_1}_\kk}+\frac{n^{\lambda_3\lambda_2}_\kk}{2\epsilon^+-\varepsilon^{\lambda_3\lambda_2}_\kk}\Biggl) + \frac{\mathcal T^{3,\alpha,\beta,\gamma,\delta}_{\lambda_1\lambda_2\lambda_3,\kk}}{3\epsilon^+ -\varepsilon^{\lambda_3\lambda_1}_{\kk}} \Biggl(\frac{n^{\lambda_1\lambda_2}_\kk}{2\epsilon^+-\varepsilon^{\lambda_2\lambda_1}_\kk}+ \frac{n^{\lambda_3\lambda_2}_\kk}{\epsilon^+-\varepsilon^{\lambda_3\lambda_2}_\kk}\Biggl)\Biggl],
\end{aligned}
\eeq
\end{widetext}
where again $\mathcal T^{i,\alpha\beta\gamma\delta}_{\{\lambda_i\},\kk}$ are the vertex functions. Akin to the bubble diagram case, the calculation of the real part can be reduced to a polynomial integral; hence, it is a moot exercise to repeat the calculations. We obtain an analog resonant structure for the real part of the $zzzz$-term at zero temperature,
\beq
\begin{split}
\text{Re} [T_{zzzz}(3\epsilon,\epsilon,\epsilon,\epsilon)] = \frac{e^4 a_z}{8 \pi \hbar\,\eta^2 \epsilon^3}\\ \Bigl[t^1_-(\epsilon,\tau)\Theta\Bigl(\hbar \epsilon-\frac{2\mu}{1+\tau}\Bigl)  + t^2_-(\epsilon,\tau) \Theta\Bigl(\hbar \epsilon-\frac{\mu}{1+\tau)}\Bigl) + \\ t^3_-(\epsilon,\tau) \Theta\Bigl(\hbar \epsilon-\frac{2\mu}{3(1+\tau)}\Bigl) + t^1_+(\epsilon,\tau)\Theta\Bigl(\hbar \epsilon-\frac{2\mu}{\tau-1}\Bigl)  + \\ t^2_+(\epsilon,\tau) \Theta\Bigl(\hbar \epsilon-\frac{\mu}{\tau-1)}\Bigl) + t^3_+(\epsilon,\tau) \Theta\Bigl(\hbar \epsilon-\frac{2\mu}{3(\tau-1)}\Bigl)\Bigl],
\end{split}
\eeq
where the $t^i_\pm(\epsilon,\tau)$ functions are defined as
\begin{widetext}
\begin{equation}
    \begin{aligned}
 t_{\mp}^{1}(\eps,\tau) &=\int_{\gamma^{(1)}_\mp} dw\,[2\tilde {\mathcal{ T}}^{1,zzzz}_{-1,1,-1} (w,\tau) + \tilde {\mathcal{ T}}^{2,zzzz}_{1,-1,1} (w,\tau)-\tilde {\mathcal{ T}}^{3,zzzz}_{-1,1,-1} (w,\tau)+\tilde{\mathcal{ T}}^{1,zzzz}_{-1,-1,1} (w,\tau)+&\\& + \tilde {\mathcal{ T}}^{2,zzzz}_{-1,-1,1} (w,\tau) +\tilde {\mathcal{  T}}^{1,zzzz}_{-1,1,1} (w,\tau) + \tilde {\mathcal{ T}}^{3,zzzz}_{-1,1,1} (w,\tau)]\\
    t_{\mp}^{2}(\eps,\tau)& = \frac{1}2\int_{\gamma^{(2)}_\mp} dw\,[-2\tilde {\mathcal{ T}}^{2,zzzz}_{-1,1,-1} (w,\tau) +2 \tilde {\mathcal{ T}}^{3,zzzz}_{1,-1,1} (w,\tau)-\tilde{\mathcal{ T}}^{1,zzzz}_{-1,-1,1} (w,\tau)+&\\& + \tilde {\mathcal{ T}}^{3,zzzz}_{-1,-1,1} (w,\tau) -\tilde {\mathcal{  T}}^{1,zzzz}_{-1,1,1} (w,\tau) + \tilde {\mathcal{ T}}^{2,zzzz}_{-1,1,1} (w,\tau)]\\
t_\mp^1(\eps,\tau)&=\int_{\gamma^{(3)}_\mp}dw\,[\tilde{\mathcal{ T}}^{2,zzzz}_{-1,-1,1} (w,\tau) + \tilde {\mathcal{ T}}^{3,zzzz}_{-1,-1,1} (w,\tau) -\tilde {\mathcal{  T}}^{2,zzzz}_{-1,1,1} (w,\tau) + \tilde {\mathcal{ T}}^{3,zzzz}_{-1,1,1} (w,\tau)]
    \end{aligned}
\end{equation}
where the $\gamma$-intervals are given by $\gamma^{(n)}_\mp=[-(2\mu-n \epsilon)/(n(\tau+1)),1], \,\,[(2\mu-n \epsilon)/(n(\tau+1)),(2\mu+n \epsilon)/(n(\tau+1))] $
and the rescaled vertex functions reads
\begin{align}
  \tilde {\mathcal{ T}}^{n,zzzz}_{1,-1,1}(w,\tau) &=\left\{\frac{1}{4} \left(1-w^2\right)^2,-2 w \left(1-w^2\right) (\tau +w),-w \left(1-w^2\right) (\tau +w)\right\},\\
    \tilde {\mathcal{ T}}^{n,zzzz}_{-1,1,-1}(w,\tau)& =\left\{\frac{1}{4} \left(1-w^2\right)^2, w \left(1-w^2\right) (\tau +w),2w \left(1-w^2\right) (\tau +w)\right\},\\
    \tilde {\mathcal{ T}}^{n,zzzz}_{-1,-1,1}(w,\tau)&=\left\{2 w \left(1-w^2\right) (w-\tau ),\frac{1}{2} \left(1-w^2\right)^2,4 w \left(1-w^2\right) (w-\tau )\right\},\\
    \tilde {\mathcal T}^{n,zzzz}_{-1,1,1}(w,\tau)&=\left\{w \left(1-w^2\right) (\tau +w),6 w \left(1-w^2\right) (w-\tau ),\frac{1}{4} \left(1-w^2\right)^2\right\},
 \end{align} 
 \end{widetext}
 where we only considered the terms contributing to the positive resonances. The calculation of the square diagram is done in the same way. 
\bibliography{Mybib}

@book{boyd,
	address = {Burlington},
	author = {Robert W. Boyd},
	doi = {https://doi.org/10.1016/B978-0-12-369470-6.00016-2},
	edition = {Third Edition},
	isbn = {978-0-12-369470-6},
	publisher = {Academic Press},
	title = {Nonlinear Optics},
	url = {http://www.sciencedirect.com/science/article/pii/B9780123694706000162},
	year = {2008}}

@article{bm,
	author = {Brown, D. J. and McPhail, A. V. H. and White, D. H. and Baillie, D. and Ruddell, S. K. and Hoogerland, M. D.},
	doi = {10.1103/PhysRevA.98.013606},
	issue = {1},
	journal = {Phys. Rev. A},
	numpages = {6},
	pages = {013606},
	publisher = {American Physical Society},
	title = {Thermalization, condensate growth, and defect formation in an out-of-equilibrium Bose gas},
	url = {https://link.aps.org/doi/10.1103/PhysRevA.98.013606},
	volume = {98},
	year = {2018}}

@article{neto,
	author = {Castro Neto, A. H. and Guinea, F. and Peres, N. M. R. and Novoselov, K. S. and Geim, A. K.},
	doi = {10.1103/RevModPhys.81.109},
	issue = {1},
	journal = {Rev. Mod. Phys.},
	numpages = {0},
	pages = {109},
	publisher = {American Physical Society},
	title = {The electronic properties of graphene},
	url = {https://link.aps.org/doi/10.1103/RevModPhys.81.109},
	volume = {81},
	year = {2009}}

@article{m,
	author = {Kin Fai Mak and Changgu Lee and James Hone and Jie Shan and Tony F. Heinz},
	journal = {Phys. Rev. Lett.},
	pages = {136805},
	title = {Atomically thin MoS2: A new direct-gap semiconductor.},
	volume = {105},
	year = {2010}}

@article{I,
	author = {K. L. Ishikawa},
	journal = {Phys. Rev. B},
	pages = {201402},
	title = {Nonlinear optical response of graphene in time domain},
	volume = {82},
	year = {2010}}

@article{ng,
	author = {K. S. Novoselov and A. K. Geim and S. V. Morozov and D. Jiang, Y. Zhang and S. V. Dubonos and I. V. Grigorieva and A. A. Firsov},
	journal = {Science},
	pages = {666},
	title = {Electric field effect in atomi- cally thin carbon films},
	volume = {306},
	year = {2004}}

@article{nb,
	author = {R. R. Nair and P. Blake and A. N. Grigorenko and K. S. Novoselov and T. J. Booth and T. Stauber and N. M. R. Peres and A. K. Geim},
	journal = {Science},
	pages = {1308},
	title = {Fine Structure Constant Defines Visual Transparency of Graphene},
	volume = {320},
	year = {2008}}

@article{wb,
	author = {T. O. Wehling and A.M. Black-Shaffer and A. V. Balatsky},
	journal = {Advances in Physics},
	pages = {1},
	title = {Dirac Materials},
	volume = {63},
	year = {2013}}

@article{ws,
	author = {Z.Wang, Y. Sun and X.-Q. Chen and C. Franchini and G. Xu, H.Weng, X.
Dai and Z. Fang},
	journal = {Phys. Rev. B},
	pages = {195320},
	title = {Dirac semimetal and topological phase transitions in \uppercase{A}$_3$\uppercase{B}i (\uppercase{A}=\uppercase{N}a, \uppercase{K}, \uppercase{R}b)},
	volume = {85},
    url={https://doi.org/10.1103/PhysRevB.85.195320},
	year = {2012}}

@article{ww,
	author = {Z. Wang and H. Weng and Q. Wu and X. Dai and Z. Fang},
	journal = {Phys. Rev. B},
	pages = {125427},
	title = {Three-dimensional Dirac semimetal and quantum transport in \uppercase{C}d$_3$\uppercase{A}s$_2$},
	volume = {88},
    url={https://doi.org/10.1103/PhysRevB.88.125427},
	year = {2013}}

@article{tsa,
	author = {A. Thakur and K. Sadhukhan and A. Agarwal},
	journal = {Phys. Rev. B},
	pages = {035403},
	title = {Dynamic current-current susceptibility in three-dimensional Dirac and Weyl semimetals},
	volume = {97},
    url={ https://doi.org/10.1103/PhysRevB.97.035403},
	year = {2018}}

@article{xl1,
	author = {Mingsheng Xu and Tao Liang and Minmin Shi and Hongzheng Chen},
	journal = {Chem. Rev.},
	pages = {3766-3798},
	title = {Graphene-Like Two-Dimensional Materials},
	volume = {113},
    url={https://doi.org/10.1021/cr300263a},
	year = {2013}}

@article{hak,
	author = {M. Z. Hasan and C. L. Kane},
	journal = {Rev. Mod. Phys},
	pages = {3045},
	title = {Colloquium: Topological insulators},
	volume = {82},
    url={https://doi.org/10.1103/RevModPhys.82.3045},
	year = {2010}}

@article{Soluyanov2015,
author={Soluyanov, Alexey A.
and Gresch, Dominik
and Wang, Zhijun
and Wu, QuanSheng
and Troyer, Matthias
and Dai, Xi
and Bernevig, B. Andrei},
title={Type-\uppercase{II} \uppercase{W}eyl semimetals},
journal={Nature},
year={2015},
volume={527},
number={7579},
url={https://doi.org/10.1038/nature15768}
}

@article{dwd,
	author = {Deng, K. and Wan, G and Deng, P. Wan, Guoliang and Deng, Peng
and Zhang, Kenan and Ding, Shijie and - Wang, Eryin and  Yan, Mingzhe and  Huang, Huaqing and  Zhang, Hongyun and  Xu, Zhilin and  Denlinger, Jonathan and  Fedorov, Alexei and  Yang, Haitao and Duan, Wenhui and  Yao, Hong and  Wu, Yang and  Fan, Shoushan and  Zhang, Haijun and  Chen, Xi and  Zhou, Shuyun },
	journal = {Nat. Phys.},
	pages = {1105–1110 },
	title = {Experimental observation of topological Fermi arcs in type-\uppercase{II} Weyl semimetal \uppercase{M}o\uppercase{T}e$_2$},
	volume = {12},
    url={https://doi.org/10.1038/nphys3871},
	year = {2016}}

@article{wyh,
  title = {Observation of \uppercase{F}ermi arc and its connection with bulk states in the candidate type-\uppercase{II} \uppercase{W}eyl semimetal ${\mathrm{WTe}}_{2}$},
  author = {Wang, Chenlu and Zhang, Yan and Huang, Jianwei and Nie, Simin and Liu, Guodong and Liang, Aiji and Zhang, Yuxiao and Shen, Bing and Liu, Jing and Hu, Cheng and Ding, Ying and Liu, Defa and Hu, Yong and He, Shaolong and Zhao, Lin and Yu, Li and Hu, Jin and Wei, Jiang and Mao, Zhiqiang and Shi, Youguo and Jia, Xiaowen and Zhang, Fengfeng and Zhang, Shenjin and Yang, Feng and Wang, Zhimin and Peng, Qinjun and Weng, Hongming and Dai, Xi and Fang, Zhong and Xu, Zuyan and Chen, Chuangtian and Zhou, X. J.},
  journal = {Phys. Rev. B},
  volume = {94},
  issue = {24},
  pages = {241119},
  numpages = {8},
  year = {2016},
  month = {Dec},
  publisher = {American Physical Society},
  url = {https://link.aps.org/doi/10.1103/PhysRevB.94.241119}
}

@article{Yan2017,
author={Yan, Mingzhe
and Huang, Huaqing
and Zhang, Kenan
and Wang, Eryin
and Yao, Wei
and Deng, Ke
and Wan, Guoliang
and Zhang, Hongyun
and Arita, Masashi
and Yang, Haitao
and Sun, Zhe
and Yao, Hong
and Wu, Yang
and Fan, Shoushan
and Duan, Wenhui
and Zhou, Shuyun},
title={Lorentz-violating type-\uppercase{II} Dirac fermions in transition metal dichalcogenide \uppercase{P}t\uppercase{T}e$_2$},
journal={Nature Communications},
year={2017},
month={Aug},
day={15},
volume={8},
number={1},
pages={257},
url={https://doi.org/10.1038/s41467-017-00280-6}
}

@article{Huang2016,
  title = {Type-\uppercase{II} \uppercase{D}irac fermions in the ${\mathbf{PtSe}}_{\mathbf{2}}$ class of transition metal dichalcogenides},
  author = {Huang, Huaqing and Zhou, Shuyun and Duan, Wenhui},
  journal = {Phys. Rev. B},
  volume = {94},
  issue = {12},
  pages = {121117},
  numpages = {6},
  year = {2016},
  month = {Sep},
  publisher = {American Physical Society},
  doi = {10.1103/PhysRevB.94.121117},
  url = {https://link.aps.org/doi/10.1103/PhysRevB.94.121117}
}

@article{Volovik2017,
author={Volovik, G. E.
and Zhang, K.},
title={Lifshitz Transitions, Type-\uppercase{II} \uppercase{D}irac and \uppercase{W}eyl Fermions, Event Horizon and All That},
journal={Journal of Low Temperature Physics},
year={2017},
volume={189},
number={5},
pages={276-299},
url={https://doi.org/10.1007/s10909-017-1817-8}
}

@article{Cheng_2014,
	author = {Cheng, J L and Vermeulen, N and Sipe, J E},
	doi = {10.1088/1367-2630/16/5/053014},
	journal = {New Journal of Physics},
	number = {5},
	pages = {053014},
	publisher = {IOP Publishing},
	title = {Third order optical nonlinearity of graphene},
	url = {https://dx.doi.org/10.1088/1367-2630/16/5/053014},
	volume = {16},
	year = {2014}}

@book{Giuliani_Vignale_2005, place={Cambridge}, title={Quantum Theory of the Electron Liquid}, publisher={Cambridge University Press}, author={Giuliani, Gabriele and Vignale, Giovanni}, year={2005}}

@article{Parker2019_DiagrammaticNonlinear,
  author       = {Parker, Daniel E. and Morimoto, Takahiro and Orenstein, Joseph and Moore, Joel E.},
  title        = {Diagrammatic approach to nonlinear optical response with application to Weyl semimetals},
  journal      = {Physical Review B},
  volume       = {99},
  number       = {045121},
  pages        = {1--15},
  year         = {2019},
  publisher    = {American Physical Society (APS)},
  doi          = {10.1103/PhysRevB.99.045121},
  url          = {https://doi.org/10.1103/PhysRevB.99.045121}
}

@article{Zhu2023Ultrastrong,
  author = {Zhu, Song and Duan, Ruihuan and Chen, Wenduo and Wang, Fakun and Han, Jiayue and Xu, Xiaodong and Wu, Lishu and Ye, Ming and Sun, Fangyuan and Song, Han and Han, Song E. and Zhao, Xiaoxu and Tan, Chuan Seng and Liang, Houkun and Liu, Zheng and Wang, Qi Jie},
  title = {Ultrastrong Optical Harmonic Generations in Layered Platinum Disulfide in the Mid-Infrared},
  journal = {ACS Nano},
  year = {2023},
  volume = {17},
  number = {3},
  pages = {2148--2158},
  doi = {10.1021/acsnano.2c08147},
  url = {https://pubs.acs.org/doi/10.1021/acsnano.2c08147},
  publisher = {American Chemical Society (ACS)}
}

@article{Carbotte2016DiracConeTilt,
  title     = {Dirac cone tilt on interband optical background of type-I and type-II Weyl semimetals},
  author    = {Carbotte, J. P.},
  journal   = {Physical Review B},
  volume    = {94},
  number    = {16},
  pages     = {165111},
  year      = {2016},
  month     = oct,
  doi       = {10.1103/PhysRevB.94.165111},
  publisher = {American Physical Society},
  url       = {https://doi.org/10.1103/PhysRevB.94.165111}
}

@article{Tan2022,
  title = {Signatures of Lifshitz transition in the optical conductivity of two-dimensional tilted Dirac materials},
  author = {Tan, Chao-Yang and Hou, Jian-Tong and Yan, Chang-Xu and Guo, Hong and Chang, Hao-Ran},
  journal = {Phys. Rev. B},
  volume = {106},
  issue = {16},
  pages = {165404},
  numpages = {22},
  year = {2022},
  month = {Oct},
  publisher = {American Physical Society},
  doi = {10.1103/PhysRevB.106.165404},
  url = {https://link.aps.org/doi/10.1103/PhysRevB.106.165404}
}

@article{Tamashevich2022PRB105_195102,
  author    = {Tamashevich, Yaraslau and Di Mauro Villari, Leone and Ornigotti, Marco},
  title     = {Nonlinear optical response of type-II Weyl fermions in two dimensions},
  journal   = {Physical Review B},
  volume    = {105},
  number    = {19},
  pages     = {195102},
  year      = {2022},
  doi       = {10.1103/PhysRevB.105.195102},
  publisher = {American Physical Society}
}

@article{Tamashevich2023PRB107_245425,
  author    = {Tamashevich, Yaraslau and Di Mauro Villari, Leone and Ornigotti, Marco},
  title     = {Two-dimensional Weyl materials in the presence of constant magnetic fields},
  journal   = {Physical Review B},
  volume    = {107},
  number    = {24},
  pages     = {245425},
  year      = {2023},
  doi       = {10.1103/PhysRevB.107.245425},
  publisher = {American Physical Society}
}

@article{DiMauroVillariPrincipi2022,
  author    = {Di Mauro Villari, Leone and Principi, Alessandro},
  title     = {Optotwistronics of bilayer graphene},
  journal   = {Physical Review B},
  volume    = {106},
  number    = {3},
  pages     = {035401},
  year      = {2022},
  doi       = {10.1103/PhysRevB.106.035401},
  publisher = {American Physical Society},
  url       = {https://doi.org/10.1103/PhysRevB.106.035401}
}

@article{Sie2019UltrafastSymmetrySwitch,
  title   = {An ultrafast symmetry switch in a Weyl semimetal},
  author  = {Sie, Edbert J. and Nyby, Clara M. and Pemmaraju, C. D. and Park, Su Ji and Shen, Xiaozhe and others},
  journal = {Nature},
  volume  = {565},
  pages   = {61--66},
  year    = {2019},
  doi     = {10.1038/s41586-018-0809-4}
}

@article{Heide2022ProbingTPT_HHG,
  title   = {Probing topological phase transitions using high-harmonic generation},
  author  = {Heide, Christian and Kobayashi, Yuki and Baykusheva, Denitsa R. and Jain, Deepti and Sobota, Jonathan A. and Hashimoto, Makoto and Kirchmann, Patrick S. and Oh, Seongshik and Heinz, Tony F. and Reis, David A. and Ghimire, Shambhu},
  journal = {Nature Photonics},
  volume  = {16},
  pages   = {620--624},
  year    = {2022},
  doi     = {10.1038/s41566-022-01050-7}
}

@article{Bauer2022OpticallySensingTPT,
  title   = {Optically sensing topological phase transitions},
  author  = {Bauer, Dieter},
  journal = {Nature Photonics},
  volume  = {16},
  pages   = {614--615},
  year    = {2022},
  doi     = {10.1038/s41566-022-01063-2}
}

@article{Morimoto2016TopologicalNLO,
  title   = {Topological nature of nonlinear optical effects in solids},
  author  = {Morimoto, Takahiro and Nagaosa, Naoto},
  journal = {Science Advances},
  volume  = {2},
  number  = {5},
  pages   = {e1501524},
  year    = {2016},
  doi     = {10.1126/sciadv.1501524}
}

@book{HaugKoch2009QuantumTheorySemiconductors,
  title     = {Quantum Theory of the Optical and Electronic Properties of Semiconductors},
  author    = {Haug, Hartmut and Koch, Stephan W.},
  edition   = {5},
  year      = {2009},
  publisher = {World Scientific Publishing},
  address   = {Singapore},
  doi       = {10.1142/7184},
  isbn      = {978-981-283-883-4}
}

@article{Giannozzi2009,
  author  = {Giannozzi, Paolo and Baroni, Stefano and Bonini, Nicola and Calandra, Matteo and Car, Roberto and Cavazzoni, Carlo and Ceresoli, Davide and Chiarotti, Gian Luca and Cococcioni, Matteo and Dabo, Ismaila and others},
  title   = {{QUANTUM ESPRESSO}: a modular and open-source software project for quantum simulations of materials},
  journal = {Journal of Physics: Condensed Matter},
  year    = {2009},
  volume  = {21},
  number  = {39},
  pages   = {395502},
  doi     = {10.1088/0953-8984/21/39/395502}
}

@article{Giannozzi2017,
  author  = {Giannozzi, Paolo and Andreussi, Oliviero and Brumme, Tobias and Bunau, Olga and Nardelli, Marco Buongiorno and Calandra, Matteo and Car, Roberto and Cavazzoni, Carlo and Ceresoli, Davide and Cococcioni, Matteo and others},
  title   = {Advanced capabilities for materials modelling with {Quantum ESPRESSO}},
  journal = {Journal of Physics: Condensed Matter},
  year    = {2017},
  volume  = {29},
  number  = {46},
  pages   = {465901},
  doi     = {10.1088/1361-648X/aa8f79}
}

@article{vanSetten2018,
  author  = {van Setten, Michiel J. and Giantomassi, Matteo and Bousquet, Eric and Verstraete, Matthieu J. and Hamann, D. R. and Gonze, Xavier and Rignanese, Gian-Marco},
  title   = {The {PseudoDojo}: Training and grading a 85 element optimized norm-conserving pseudopotential table},
  journal = {Computer Physics Communications},
  year    = {2018},
  volume  = {226},
  pages   = {39--54},
  doi     = {10.1016/j.cpc.2018.01.012},
  url     = {http://www.pseudo-dojo.org/}
}

@article{Perdew1996,
  author  = {Perdew, John P. and Burke, Kieron and Ernzerhof, Matthias},
  title   = {Generalized Gradient Approximation Made Simple},
  journal = {Physical Review Letters},
  year    = {1996},
  volume  = {77},
  number  = {18},
  pages   = {3865--3868},
  doi     = {10.1103/PhysRevLett.77.3865}
}

@article{Monkhorst1976,
  author  = {Monkhorst, Hendrik J. and Pack, James D.},
  title   = {Special points for {Brillouin}-zone integrations},
  journal = {Physical Review B},
  year    = {1976},
  volume  = {13},
  number  = {12},
  pages   = {5188--5192},
  doi     = {10.1103/PhysRevB.13.5188}
}

@article{Ikeda2020,
  author    = {Ikeda, Tatsuhiko N.},
  title     = {High-order nonlinear optical response of a twisted bilayer graphene},
  journal   = {Phys. Rev. Research},
  volume    = {2},
  number    = {3},
  pages     = {032015},
  year      = {2020},
  doi       = {10.1103/PhysRevResearch.2.032015},
  publisher = {American Physical Society}
}

@article{ZuberZhang2021,
  author    = {Zuber, J. W. and Zhang, C.},
  title     = {Nonlinear optical response of twisted bilayer graphene},
  journal   = {Phys. Rev. B},
  volume    = {103},
  number    = {24},
  pages     = {245417},
  year      = {2021},
  doi       = {10.1103/PhysRevB.103.245417},
  publisher = {American Physical Society}
}

@article{CarvalhoMariniBiancalana2018,
  author    = {Carvalho, David N. and Biancalana, Fabio and Marini, Andrea},
  title     = {Nonlinear optical effects of opening a gap in graphene},
  journal   = {Phys. Rev. B},
  volume    = {97},
  number    = {19},
  pages     = {195123},
  year      = {2018},
  doi       = {10.1103/PhysRevB.97.195123},
  publisher = {American Physical Society}
}

@article{DiPietroMitrano2018,
  title        = {Emergent {Dirac} carriers across a pressure-induced {Lifshitz} transition in black phosphorus},
  author       = {Di Pietro, P. and Mitrano, M. and Caramazza, S. and Capitani, F. and Lupi, S. and Postorino, P. and Ripanti, F. and Joseph, B. and Ehlen, N. and Gr{\"u}neis, A. and Sanna, A. and Profeta, G. and Dore, P. and Perucchi, A.},
  journal      = {Phys. Rev. B},
  volume       = {98},
  number       = {16},
  pages        = {165111},
  year         = {2018},
  month        = oct,
  publisher    = {American Physical Society},
  doi          = {10.1103/PhysRevB.98.165111},
  url          = {https://link.aps.org/doi/10.1103/PhysRevB.98.165111}
}

@article{Macis2024,
  title        = {Terahertz and Infrared Plasmon Polaritons in {PtTe}$_2$ Type-II Dirac Topological Semimetal},
  author       = {Macis, Salvatore and D'Arco, Annalisa and Mosesso, Lorenzo and Paolozzi, Maria Chiara and Tofani, Silvia and Tomarchio, Luca and Tummala, Pinaka Pani and Ghomi, Sara and Stopponi, Veronica and Bonaventura, Eleonora and Massetti, Chiara and Codegoni, Davide and Serafini, Andrea and Targa, Paolo and Zacchigna, Michele and Lamperti, Alessio and Martella, Christian and Molle, Alessandro and Lupi, Stefano},
  journal      = {Advanced Materials},
  year         = {2024},
  volume       = {36},
  number       = {29},
  pages        = {2400554},
  doi          = {10.1002/adma.202400554},
  url          = {https://doi.org/10.1002/adma.202400554},
  note         = {Article e2400554}
}
\end{document}